\documentclass[10pt,a4paper]{article}

\usepackage{graphicx}
\usepackage{epsfig}
\usepackage{amsfonts}
\usepackage{amssymb}
\usepackage{amsthm}
\usepackage{amsmath}
\usepackage{hyperref}
\usepackage{mathbbol}
\usepackage{multicol}
\numberwithin{equation}{section}
\newcommand{\squarecell}[9]{
	\xymatrix@R=1pc@C=5pc{
		#1 \ar@{->}[r]|{#2}="2" \ar[d]_{#4} & #3 \ar[d]^{#6} \\
		#7 \ar@{->}[r]|{#8}="8" & #9  \\
		\ar@{=>}^{#5} "2";"8"}}

\begin{document}

\setcounter{equation}{0}
\setcounter{section}{0}

\begin{center}
{\Large \bf Electro-Magnetic Space-Time Duality for\\ $\bf 2+1$-Dimensional Stationary Classical Solutions\\[5mm]}

\noindent {\noindent \bf P. Castelo Ferreira\\[5mm]}
\noindent Centre for Rapid and Sustainable Product Development\\
Polytechnic Institute of Leiria\\[5mm]
pedro.castelo.ferreira@gmail.com
\end{center}

\paragraph{Abstract}
In this paper it is studied a space-time duality web that maps electric into magnetic (and magnetic into electric) charged classical stationary rotating solutions for $2+1$-dimensional Abelian Einstein Maxwell Chern-Simons theories. A first duality map originally suggested by Kogan for static charged solutions is extended to stationary rotating space-times and are suggested two new space-time dualities maps. The three dualities complete a close duality web. It is also shown that in $3+1$-dimensions these dualities are only possible for systems which exhibit non-projected cylindrical symmetry and are not related to the standard electromagnetic duality of Maxwell equations which acts on the physical fields and charges. Generalization to $N$-form theories in higher dimensional space-times is briefly discussed.


\section{Introduction~\label{sec.intro}}

In an attempt to justify the quantization of electric charge $e$ Dirac concluded that magnetic charge $g$ should also exist in nature and it is consistently predicted from the quantum theory solutions~\cite{Dirac}. Although magnetic monopoles have never been detected experimentally this theoretical observation is still today the best justification for the experimentally observed quantization of electric charge. Following Dirac argument a generalization of the standard Maxwell electromagnetism explicitly including magnetic charges is straight forward achievable by considering both electric 4-currents $J_e = (\rho_e,{\bf j_e})$ and magnetic 4-currents $J_g = (\rho_g,{\bf j_g})$, where the $\rho$'s are the charge densities and the ${\bf j}$'s are the current densities. In particular was noticed that the covariant $3+1$-dimensional Maxwell equations, with both electric and magnetic currents, are invariant under the following rotation of the electromagnetic fields ${\bf E} + {\bf iB}\,\leftrightarrow\, e^{i\phi}({\bf E} + {\bf iB})$ and $J_e + iJ_g\,\leftrightarrow\, e^{i\phi}(J_e + iJ_g)$, where ${\bf E}$ is the electric field and ${\bf B}$ is the magnetic field (in this paper are employing natural units $\hbar=c=1$). This field map is generally known as the electromagnetic duality (see~\cite{Jackson,Olive} for a review in the subject).

The electromagnetic duality was further explored in the context of $3+1$-dimensional charged gravitational solutions, namely in the works of Deser and al.~\cite{Deser_01, Deser_02} this field duality is applied to black hole solutions in several dimensions, such that computed electric solutions are mapped into dual magnetic solutions. Also in $2+1$-dimensional (planar) systems exact field and gravitational electric and magnetic charged solutions have been extensively studied in the literature~\cite{Kogan1,Kogan2,cyl_01, cyl_02, Stichel, DeserTekin, bh_CSmass, electric, oscar} (see also references therein). The motivation for these works is not purely theoretical as $2+1$-dimensional systems effectively describe real physical phenomena in condensed matter systems~\cite{Wilczek} and experimental searches for both elementary magnetic monopoles as well as for extended magnetically charged solutions in $3+1$- and $2+1$-dimensional systems are justified by the original Dirac argument~\cite{monopoles}.

For planar systems, a duality between electric and magnetic charged solutions was suggested by Kogan~\cite{Kogan1,Kogan2}. This duality acts directly in the $2+1$-dimensional space-time manifold ($t\to i\varphi$ and $\varphi\to it$) instead of acting in the electromagnetic fields and was applied to Abelian Einstein Maxwell Chern-Simons theories for static radial symmetric metrics. In this paper this map is extended to stationary rotating spaces and a new distinct space-time duality map is suggested ($t\to \varphi$ and $\varphi\to t$) that also maps electric into magnetic solutions (and magnetic into electric solutions). These two duality maps will further be related through a third duality map consisting of a double Wick rotation ($t\to it$ and $\varphi\to i\varphi$) such that a close duality web is obtained. The third duality does not map electric into magnetic solutions, instead maps standard gauge fields into ghost fields by swapping the relative sign between the gravitational and gauge sector of the theory. The lifting of these duality maps to $3+1$-dimensional space-time as well as their possible application to $N$-form theories in higher dimensional space-times~\cite{HigherD} is briefly discussed.

In section~\ref{sec.orig} the original duality map as suggested by Kogan is reviewed and generalized to stationary rotating space-times.
In section~\ref{sec.dual2} a new duality map is suggested and in section~\ref{sec.Wick} both dualities are related through
a third duality map which is a double Wick rotation such that a close duality web is obtained. In section~\ref{sec.4d}
the dualities are generalized to non-projected four dimensional solutions with cylindrical symmetry and their application
to $N$-form theories in higher dimensional space-times is briefly discussed.

\setcounter{equation}{0}
\section{Original Duality for $2+1$-Dimensional Rotating Space-Times\label{sec.orig}}

The action for Abelian Einstein Maxwell Chern-Simons theory in a $2+1$-dimensional Minkowski space-time manifold $M$ is
\begin{equation}
S = \int_M \left(\tilde{R}*1 - \tilde{F}\wedge*\tilde{F} + m\tilde{A}\wedge \tilde{F}\right)\ ,
\label{action}
\end{equation}
where $\tilde{R}*1$ is the Einstein term (Ricci scalar), $\tilde{F} = d\tilde{A}$ is the Maxwell tensor such that $\tilde{F}\wedge*\tilde{F}$ is the Maxwell term, $m\tilde{A}\wedge \tilde{F}$ is the topological Chern-Simons term and $m$ is the topological mass of the gauge field $\tilde{A}$ which is physically interpreted as the gauge photon field in planar systems. We briefly recall that in $2+1$-dimensional Maxwell theory the topological Chern-Simons term arises as a quantum correction and has the effect of generating a mass $m$ to the photon field $\tilde{A}$~\cite{CSmass}. Hence for consistence of the theory the Chern-Simons should be considered in the theory and does not depend explicitly on the geometry of the manifold, instead is coupled by the equations of motion for the gauge field $\tilde{A}$ to the Maxwell term, which in turn couples to the gravitational sector. Generally these theories are also known as topologically massive gauge theories. It is also relevant to further note that to avoid the gauge field being interpreted as a tachyon (having classical imaginary mass) the Maxwell and Chern-Simons term must have opposite sign (for $m>0$)~\cite{bh_CSmass}. As for the relative sign between the Einstein term and the Maxwell term imposes whether the gauge field $\tilde{A}$ is interpreted as a standard gauge field (opposite sign) or a ghost field (same sign). Specifically for standard gauge fields the quantum states have positive energy eigenvalues while for ghost fields have negative energy eigenstates (see~\cite{ACdual} and references therein for a discussion of relative sign choices and derivation of Hamiltonian quantization in $2+1$- and $3+1$-dimensional embeddings).

The original duality map consist in mapping electric into magnetic solutions (as well as magnetic into electric solutions) by exchanging the role of the time coordinate $t$ and the angular coordinate $\varphi$~\cite{Kogan1}
\begin{equation}
\begin{array}{rcl}
t&\to&i\varphi\ ,\\[5mm]
\varphi&\to&it\ .
\end{array}
\label{duality1}
\end{equation}

With the objective of generalizing this duality map to stationary rotating space-times let us
consider a $2+1$-dimensional metric on the ADM form~\cite{ADM}
\begin{equation}
ds^2=-\tilde{f}^2dt^2+dr^2+\tilde{h}^2(d\varphi+\tilde{N}dt)^2\ ,
\label{ds}
\end{equation}
and the standard electric and magnetic field definitions
\begin{equation}
\begin{array}{rcl}
\tilde{E}&=&\tilde{F}_{tr}\ ,\\[3mm]
\tilde{B}&=&\tilde{F}_{r\varphi}\ ,
\end{array}
\end{equation}
where we recall that in planar systems the magnetic field $\tilde{B}$ is a scalar while the electric field $\tilde{E}$ is a 2-vector.
Let us further introduce a Cartan triad for the metric~(\ref{ds}) $e^0=f\,dt$ and $e^2=h(d\varphi+Adt)$ (see~\cite{Carlip2}).
Then the duality map~(\ref{duality1}) is simply interpreted as the duality map for the triad elements $e^0\leftrightarrow e^2$ (see~\cite{electric}). With respect to the several gravitational fields this accounts for the following field maps
\begin{equation}
\begin{array}{rcl}
\tilde{f}&\to& ih\ ,\\[5mm]
\tilde{h}&\to& if\ ,\\[5mm]
\tilde{E}&\to& -iB\ ,\\[5mm]
\tilde{B}&\to& -iE\ ,\\[5mm]
\end{array}
\label{fieldmap1}
\end{equation}
such that the following dual metric parameterization is
\begin{equation}
ds^2_{\mathrm{dual}}=-f^2(dt+Nd\varphi)^2+dr^2+h^2d\varphi^2\ .
\label{dsB}
\end{equation}

As for the gauge sector the Maxwell and Chern-Simons terms transform as $-\tilde{F}\wedge*\tilde{F}+m\tilde{A}\wedge \tilde{F}\to+F\wedge*F-mA\wedge F$. Hence the gauge sector swaps the relative sign with respect to the gravitational sector such that standard gauge fields are mapped into ghost gauge fields.

The tensor components of the two metrics corresponding to $ds^2$ and $ds^2_{\mathrm{dual}}$ are
\begin{equation}
\left\{
\begin{array}{rcl}
\tilde{g}_{00}&=&-\tilde{f}^2+\tilde{h}^2\tilde{N}^2\\
\tilde{g}_{11}&=&1\\
\tilde{g}_{22}&=&\tilde{h}^2\\
\tilde{g}_{02}&=&\tilde{h}^2\tilde{N}
\end{array}
\right.\ , \  
\left\{
\begin{array}{rcl}
g_{00}&=&-f^2\\
g_{11}&=&1\\
g_{22}&=&h^2-f^2N^2\\
g_{02}&=&-f^2N
\end{array}\right.\ ,
\end{equation}
such that these two parameterizations are straight forward related by the following equalities
\begin{equation}
\begin{array}{rcl}
\tilde{f}^2&=&\displaystyle \frac{f^2\,h^2}{h^2-f^2\,N^2}\ ,\\[6mm]
\tilde{h}^2&=&\displaystyle h^2-f^2\,N^2\ ,\\[4mm]
\tilde{N}&=&\displaystyle -\frac{N\,f^2}{h^2-f^2\,N^2}\ .
\end{array}
\label{map1}
\end{equation}

As for the ADM metric signature behavior for each specific solution, under the duality maps, are obtained the following cases
\begin{equation}
\begin{array}{lcl}
h^2-f^2\,N^2>0&\Rightarrow&\mathrm{duality\ maintains\ signature,}\\[5mm]
h^2-f^2\,N^2<0&\Rightarrow&\mathrm{duality\ changes\ signature.}
\end{array}
\label{mapsign1}
\end{equation}
This is directly concluded by inspection of the equalities~(\ref{map1}) such that the duality map~(\ref{duality1}) only
maintains the metric signature for regions of space-time where the solutions obey the constraint $h^2-f^2\,N^2>0$. This feature only arises for rotating space-times, in the limit $N\to 0$ the constraint becomes $h^2>0$ (which is verified as long as $h$ is real) such that the metric always maintains the signature under this duality.

It is relevant to stress once more that the duality~(\ref{duality1}) is a map of the space-time coordinates, although the electromagnetic fields transform accordingly~(\ref{fieldmap1}) this is not a duality map of the electromagnetic fields. Also, by swapping the relative sign between the gravitational and gauge terms, it has the effect of mapping standard gauge fields into ghost fields.

Next we will show that this duality map is not the only possible duality.

\setcounter{equation}{0}
\section{Another Possible Duality for Rotating Space-Times\label{sec.dual2}}

Generally, for a given particular solution already computed, it may be desired to obtain the opposite sign transformation properties for the metric signature to the one expressed by inequalities~(\ref{mapsign1}). Hence it is straight forward to conclude that a possible duality map is to consider a direct swapping of the time and angular coordinates without the imaginary phase
\begin{equation}
\begin{array}{rcl}
t&\to&\varphi\ ,\\[5mm]
\varphi&\to&t\ .
\end{array}
\label{duality2}
\end{equation}
With respect to the several fields this accounts for the following field map
\begin{equation}
\begin{array}{rcl}
\tilde{f}&\to& \hat{h}\ ,\\[5mm]
\tilde{h}&\to& \hat{f}\ ,\\[5mm]
\tilde{E}&\to& -\hat{B}\ ,\\[5mm]
\tilde{B}&\to& -\hat{E}\ ,\\[5mm]
\end{array}
\label{fieldmap2}
\end{equation}
where hated fields are employed to distinguish between the two dualities given in~(\ref{duality1})
and~(\ref{duality2}). The dual metric parameterization is straight forwardly written as
\begin{equation}
d\hat{s}^2_{\mathrm{dual}}=\hat{f}^2(dt+\hat{N}d\varphi)^2+dr^2-\hat{h}^2d\varphi^2\ ,
\label{dsB2}
\end{equation}
and the gauge sector maintains its relative sign with respect to the gravitational sector as the Maxwell and Chern-Simons terms
transform as $-\tilde{F}\wedge*\tilde{F}+m\tilde{A}\wedge\tilde{F}\to-\hat{F}\wedge*\hat{F}+m\hat{A}\wedge \hat{F}$ such that standard gauge fields are mapped into standard gauge fields.

The tensor components of the dual metric~(\ref{dsB2}) are
\begin{equation}
\left\{
\begin{array}{rcl}
\hat{g}_{00}&=&\hat{f}^2\\
\hat{g}_{11}&=&1\\
\hat{g}_{22}&=&-\hat{h}^2+\hat{f}^2\hat{N}^2\\
\hat{g}_{02}&=&\hat{f}^2\hat{N}\ .
\end{array}
\right.
\end{equation}
We can relate these two parameterization by the following equality
\begin{equation}
\begin{array}{rcl}
\tilde{f}^2&=&\displaystyle \frac{\hat{f}^2\,\hat{h}^2}{-\hat{h}^2+\hat{f}^2\,\hat{N}^2}\\[6mm]
\tilde{h}^2&=&\displaystyle -\hat{h}^2+\hat{f}^2\,\hat{N}^2\\[4mm]
\tilde{N}&=&\displaystyle -\frac{\hat{N}\,\hat{f}^2}{-\hat{h}^2+\hat{f}^2\,\hat{N}^2}
\end{array}
\label{map2}
\end{equation}

As for the behavior of the metric signature under the above map, by inspection of the equalities~(\ref{map2}), we obtain the following cases:
\begin{equation}
\begin{array}{lcl}
-\hat{h}^2+\hat{f}^2\,\hat{N}^2>0&\Rightarrow&\mathrm{duality\ maintains\ signature,}\\[5mm]
-\hat{h}^2+\hat{f}^2\,\hat{N}^2<0&\Rightarrow&\mathrm{suality\ changes\ signature,}
\end{array}
\label{mapsign2}
\end{equation}
hence with the opposite sign of inequalities~(\ref{mapsign1}).

Hence, for a given solution, both the metric signature and the relative sign between the gravitational and gauge sector are mapped distinctly under the duality maps~(\ref{duality1}) and~(\ref{duality1}).

\section{Double Wick Rotation as a Duality\label{sec.Wick}}

Let us note that both dualities as given by~(\ref{duality1}) and~(\ref{duality2})
are straight forwardly related by a double Wick rotation map
\begin{equation}
\begin{array}{rcl}
t&\to& it\\[5mm]
\varphi&\to& i\varphi
\end{array}
\label{Wick}
\end{equation}
for which the fields map accordingly has
\begin{equation}
\begin{array}{rcl}
f&\to& i\hat{f}\\[5mm]
h&\to& i\hat{h}\\[5mm]
E&\to& i\hat{E}\\[5mm]
B&\to& i\hat{B}
\end{array}
\label{Wick_fields}
\end{equation}
such that the factor $-h^2+f^2\,N^2\to \hat{h}^2-\hat{f}^2\,\hat{N}^2$ swaps sign and the the gauge sector changes its relative sign with respect to the gravitational sector $+F\wedge*F-m A\wedge F\to-\hat{F}\wedge*\hat{F}+m\hat{A}\wedge \hat{F}$. Hence gauge fields are mapped into ghost gauge fields.

In this way from a given electric or magnetic solution for the metric parameterization~(\ref{ds}) one
can find a magnetic or electric solution (respectively) employing the above duality maps~(\ref{duality1})
and~(\ref{duality2}). The choice of the duality to be employed depends on the specific form of the
solutions such that the dual metric signature is set accordingly either by condition~(\ref{mapsign1})
or~(\ref{mapsign2}) and whether the gauge fields are intended to be ghost fields or standard fields.
Also by employing~(\ref{Wick}) one can, from a given charged solution, obtain another charged solution of the same kind with the
opposite ADM metric signature and distinct gauge field interpretation (for standard fields are obtained ghost fields and for ghost fields are obtained standard fields). This web of dualities is pictured in figure~\ref{fig}.

\begin{center}
	\vspace{10pt}
	\includegraphics[width=0.99\linewidth]{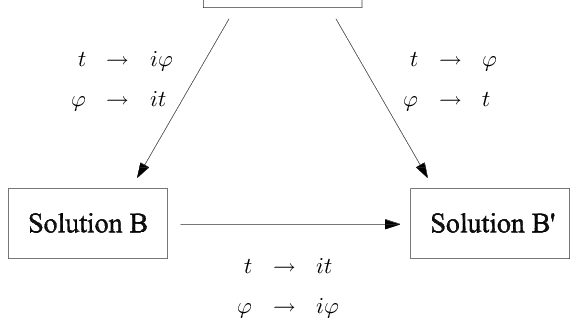}
	\par{\it Figure 1:  Web of Dualities. The dualities can be considered in both directions. In the picture
		are expressed only the dualities for the directions of the arrows. \label{fig}}
\end{center}

For last let us note that, as expected from covariance, for all the three dualities the Einstein term in the action~(\ref{action}) does not change sign as it is explicitly written in powers of 4 with respect to the metric fields $f$'s and $h$'s ($i^4 = +1$). However this fact does not imply that it is constant across space-time, generally and as expected it changes across space-time for each particular solution depending on the matter distribution.

\setcounter{equation}{0}
\section{Dualities in $3+1$D and Higher Dimensional Space-Times\label{sec.4d}}

In $2+1$-dimensions the magnetic field is a scalar (corresponding to $\tilde{B}_{(3D)}=\tilde{F}_{r\varphi}$) and
planar system can be interpreted either as a projected $3+1$-dimensional system or as embedded systems~\cite{ACdual}.
So far the dualities~(\ref{duality1}),~(\ref{duality2}) and~(\ref{Wick}) apply to both these frameworks and implicitly the angular
component of the electric field is necessarily null, $\tilde{E}_{(3D)}^\varphi=\tilde{F}^{0\varphi}=0$. This result is imposed by the Einstein equations (gravitational equation of motion) as the radial shift function in the metric~\cite{electric} does not constitute a physical gravitational degree of freedom in $2+1$-dimensional gravity~\cite{Carlip2}.

More generally in a full $3+1$-dimensional one can assume non-projected cylindrical symmetry around the $\theta$ direction
such that the projected magnetic field corresponds indeed to the $\theta$ component of the non-projected
magnetic field, $\tilde{B}_{(3D)}=\tilde{B}_{(4D)}^{\theta}$ and $\tilde{E}_{(4D)}^\varphi=0$. Concerning the projection of four dimensional
gravity to three dimensional space-times we refer the reader to~\cite{cyl_01,cyl_02}.

Then for $3+1$-dimensional solutions with cylindrical symmetry we
obtain two possible components for the electric and magnetic field
\begin{equation}
\left\{\begin{array}{rcl}
\tilde{E}_{(4D)}^r&=&\tilde{F}^{0r}\\[5mm]
\tilde{E}_{(4D)}^\theta&=&\tilde{F}^{0\theta}
\end{array}\right.
\ ,\  
\left\{\begin{array}{rcl}
\tilde{B}_{(4D)}^r&=&\tilde{F}_{\varphi\theta}\\[5mm]
\tilde{B}_{(4D)}^\theta&=&\tilde{F}_{r\varphi}
\end{array}\right.\ ,
\end{equation}
such that the dualities discussed so far map the radial and polar electric fields into the polar and radial
magnetic fields, respectively. The first field map~(\ref{fieldmap1}) is equivalent to the field map
\begin{equation}
\begin{array}{rcl}
\tilde{E}_{(4D)}^r&\to&-iB_{(4D)}^\theta\ ,\\[5mm]
\tilde{E}_{(4D)}^\theta&\to&-iB_{(4D)}^r\ ,\\[5mm]
\tilde{B}_{(4D)}^r&\to&-iE_{(4D)}^\theta\ ,\\[5mm]
\tilde{B}_{(4D)}^\theta&\to&-iE_{(4D)}^r\ ,
\end{array}
\end{equation}
and the second field map~(\ref{fieldmap2}) is equivalent to the field map
\begin{equation}
\begin{array}{rcl}
\tilde{E}_{(4D)}^r&\to&-\hat{B}_{(4D)}^\theta\ ,\\[5mm]
\tilde{E}_{(4D)}^\theta&\to&-\hat{B}_{(4D)}^r\ ,\\[5mm]
\tilde{B}_{(4D)}^r&\to&-\hat{E}_{(4D)}^\theta\ ,\\[5mm]
\tilde{B}_{(4D)}^\theta&\to&-\hat{E}_{(4D)}^r\ .
\end{array}
\end{equation}
In this way the dualities are lifted from a $2+1$-dimensional system to a $3+1$-dimensional system with cylindrical
symmetry. As expected it is not possible to generalize the dualities for generic stationary solutions to generic $3+1$-dimensional system that do not possess cylindrical symmetry. In order to conclude it, it is enough to consider the remaining field
components $\tilde{E}_{(4D)}^\varphi=\tilde{F}^{0\varphi}$ and $\tilde{B}_{(4D)}^\varphi=-\tilde{F}_{r\theta}$
which are maintained (up to sign changes) by the field maps~(\ref{fieldmap1}) and~(\ref{fieldmap2}).

It is interesting to note that a generalization is possible for $N$-form theories in higher dimensional
space-times~\cite{Deser_02}. For instance in $5+1$-dimensions, by considering the 2-form fields
$E^{IJ}=F^{0IJ}$ and $B^{IJ}=\epsilon^{IJKLM}F_{KLM}/6$, under the duality $(t\to ix^5,\ x^5\to it)$
we obtain the map $E^{12}\leftrightarrow -iB^{34}$, $E^{13}\leftrightarrow -iB^{24}$,
$E^{14}\leftrightarrow -iB^{23}$, $E^{23}\leftrightarrow -iB^{14}$, $E^{24}\leftrightarrow -iB^{13}$
and $E^{34}\leftrightarrow -iB^{12}$.

We recall once more that there is no relation of these dualities with the usual electromagnetic duality of the gauge fields~\cite{Jackson,Olive,Deser_01,Deser_02}. The original electromagnetic duality rotates the same components of the electric and magnetic into each other not mapping the space-time components of the fields as opposed to the dualities just discussed.

\section{Conclusions}

In this paper was developed a space-time duality web that allows to map electric into magnetic classical solutions into each other.
These dualities may generally change the ADM metric signature as given in~(\ref{ds}).
Then, given that the coefficient of $dr^2$ is positive (see~\cite{electric} for a discussion
on different choices of signatures), the original Minkowski signature is considered to be
the one that holds the coefficient of $dt^2$ negative and the coefficient of $d\varphi^2$ positive,
this means that we are aiming at solutions with $\tilde{f}^2>0$ and $\tilde{h}^2>0$. Also from the explicit
form of the three possible field maps~(\ref{fieldmap1}),~(\ref{fieldmap2}) and~(\ref{Wick_fields})
one may map real solutions into imaginary solutions. Nevertheless by imposing reality conditions
and properly constraining the parameters and variables of the original solutions it is generally possible to achieve real solutions.

Also the mapping of the relative sign between the gravitational sector and the gauge sector was analised such that duality~(\ref{duality1}) and~(\ref{Wick}) maps standard gauge fields and ghost gauge fields into each other while duality~(\ref{duality2}) does not.

We resume the results obtained in the next table.
\begin{equation}
\begin{array}{l|cc}
\mathrm{duality}&\mathrm{maintains\ signature}&\mathrm{standard} \leftrightarrow \mathrm{ghost}\\\hline\hline\\[-3mm]
\begin{array}{l}t\to i\varphi\\\varphi\to it\end{array}&h^2-f^2\,N^2>0&yes\\\hline\\[-3mm]
\begin{array}{l}t\to \varphi\\\varphi\to t\end{array}&h^2-f^2\,N^2<0&no\\\hline\\[-3mm]
\begin{array}{l}t\to it\\\varphi\to i\varphi\end{array}&no&yes\\\hline
\end{array}
\nonumber
\end{equation}

It is also explicitly concluded that the dualities studied are not related to the electromagnetic duality of the electromagnetic fields
and, although valid for four dimensional solutions with cylindrical symmetry, cannot be generalized to
generic solutions on $3+1$-dimensional systems. However we give an example in six dimensional space-time
that show that the dualities can be generalized to $N$-form theories in higher dimensional space-times.

\ \\
{\large\bf Acknowledgements}
This work was supported by SFRH/BPD/17683/2004 up to 2007 and by CENTRO-01-0145-FEDER-000014 from 2017 onward.


\end{document}